\documentclass[%
 reprint,
superscriptaddress,
 amsmath,amssymb,
 aps,
]{revtex4-1}

\usepackage{graphicx}
\usepackage{dcolumn}
\usepackage{bm}
\usepackage{hyperref}
\usepackage{lineno}

\begin{document}


\title{Double freeform illumination design for prescribed wavefronts and irradiances}

\author{Christoph B\"osel}
 \email{christoph.boesel@uni-jena.de}

\affiliation{Friedrich-Schiller-Universit\"at Jena, Institute of Applied Physics, Albert-Einstein-Str. 15, 07745 Jena, Germany}%
\author{Herbert Gross}%
\affiliation{Friedrich-Schiller-Universit\"at Jena, Institute of Applied Physics, Albert-Einstein-Str. 15, 07745 Jena, Germany}%
\affiliation{Fraunhofer-Institut f\"ur Angewandte Optik und Feinmechanik, Albert-Einstein-Str. 7, 07745 Jena, Germany}

\begin{abstract}
A mathematical model in terms of partial differential equations (PDE) for the calculation of double freeform
surfaces for irradiance and phase control with predefined input and output wavefronts is presented. It extends
the results of B\"osel and Gross [J. Opt. Soc. Am. A 34, 1490 (2017)] for the illumination design of single freeform
surfaces for zero-étendue light sources to double freeform lenses and mirrors. The PDE model thereby overcomes
the restriction to paraxiality or the requirement of at least one planar wavefront of the current design models in
the literature. In contrast with the single freeform illumination design, the PDE system does not reduce to a
Monge–Ampère type equation for the unknown freeform surfaces, if nonplanar input and output wavefronts
are assumed. Additionally, a numerical solving strategy for the PDE model is presented. To show its efficiency,
the algorithm is applied to the design of a double freeform mirror system and double freeform lens system.

\url{https://doi.org/10.1364/JOSAA.35.000236}
\end{abstract}

\maketitle



\section{Introduction}
\label{sec:1} 

Whereas there have been proposed numerous numerical design algorithms for illumination control with single freeform surfaces in recent years, design methods for irradiance \textit{and} phase control for systems without symmetries are rather rare \cite{Ries_05, Feng13_2, Feng17_1, Wu14_1, Wu16_1, Boe17_1}. For the latter design goal, further complications arise due to the necessity of additional degrees of freedom for the phase control, which are realized by the coupling of two freeform surfaces. The design process, therefore, requires not only the consideration of the law of refraction/reflection and energy conservation law, but also the constant optical path length (OPL) condition.

One of the first attempts to solve the double freeform illumination design (DFD) problem was presented in \cite{Ries_05} by Ries, which demonstrated the design of a two off-axis mirror system for the conversion of a collimated Gaussian beam into a flat top distribution. Unfortunately, details about the design method were not given.

Later, Feng et al. presented a double freeform design algorithm \cite{Feng13_2, Feng17_1} for prescribed irradiances and wavefronts. The method is based on the calculation of a ray mapping from optimal mass transport (OMT) with a quadratic cost function and the subsequent construction of the freeform surfaces according to the mapping. Recently, a modified version of the design method was published \cite{Feng17_1}, in which the authors demonstrate the calculation of a single lens for mapping two target irradiances onto seperated planes. The design method is able to handle also complicated irradiance boundaries, but is inherently restricted by the quadratic cost function mapping to paraxiality and a thin-lens approximation, respectively. This connection between a specific design problem and an appropriate cost function was discussed in several publications (e.g. \cite{Glimm04_2, Rub07_1, Glimm10_3, Oliker11_2}). 

To overcome the restriction to paraxiality, Bösel and Gross \cite{Boe17_1} calculated an appropriate ray mapping for the collimated beam shaping with two freeform surfaces by solving a system of coupled PDE's with the quadratic cost function mapping as an intial iterate. The presented method was thereby restricted to planar input and output wavefronts.

As an alternative approach to calculate two freeform surfaces for irradiance and phase control, the DFD problem was formulated by Zhang et al. \cite{Wu14_1} and Chang et al. \cite{Wu16_1} in terms of a MAE. Subsequently, the PDE was solved by using finite differences and applying the Newton method to find a root of the resulting nonlinear equation system. The restriction of the presented design method is thereby the necessity of at least one planar wavefront.

Since the proposed numerical methods for irradiance and phase control in literature are either inherently restricted to paraxiality \cite{Feng13_2, Feng17_1} or not able to handle arbitrary input \textit{and} output wavefronts \cite{Ries_05, Wu14_1, Wu16_1, Boe17_1}, we present a mathematical model without these restrictions. The model is thereby an extension of our work on SFD  for prescribed input wavefronts of zero-\'etendue light sources  \cite{Boe17_2}. In \cite{Boe17_2}, a PDE system and MAE, respectively, for the unknown freeform surface and corresponding ray-mapping was derived. This allowed the calculation of freeform surfaces for nonspherical and nonplanar input wavefronts, which occur e.g. if the wavefront of a  point source is deflected by a reflective or refractive surface.

As we will demonstrate in section \ref{sec:2}, a similar PDE system can be derived for double freeform mirrors and lenses from the laws of refraction/reflection and the constant OPL condition. In contrast to the SFD it will turn out that the PDE system does not reduce to a MAE for the unknown freeform surfaces, if the prescribed input and output wavefront are \textit{both} nonplanar. We then propose a numerical solving strategy for the PDE system, which is implemented for square irradiance distributions. Compared to the SFD algorithm, the PDE system is thereby not solved simultaneously for the surface and the ray mapping on the \textit{target plane}. Instead, the surface and the projection of the mapping onto the output wavefront are considered. In section \ref{sec:3}, the efficiency of the design method is demonstrated for a double freeform mirror and a double freeform lens.

\section{Mathematical Model}
\label{sec:2} 

\subsection{Basic Approach}
\label{sec:2.1} 

\begin{figure}[!htb]
\begin{center}
\begin{tabular}{c}
\includegraphics[width=\linewidth]{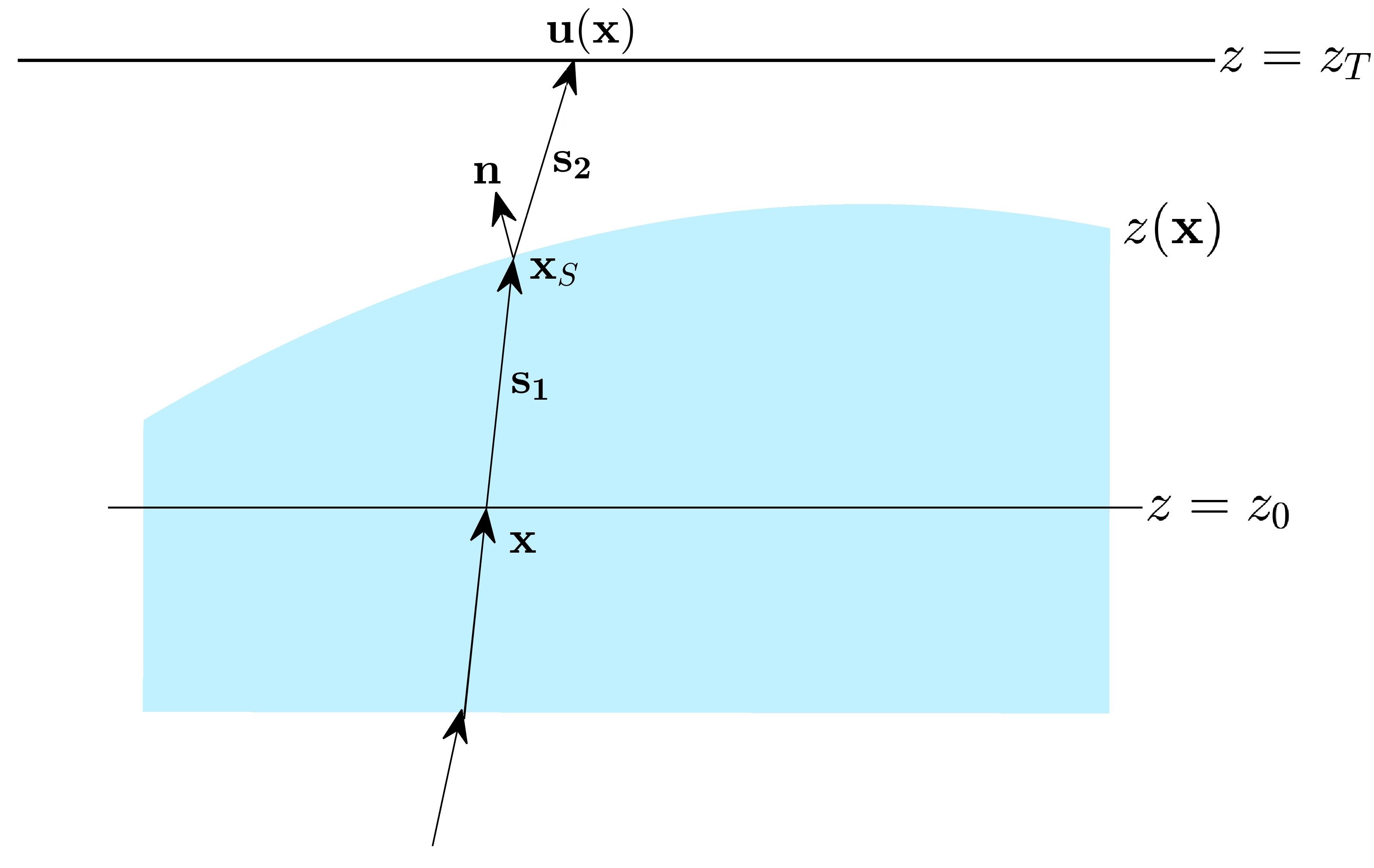}
\end{tabular}
\end{center}
\caption 
{The surface $z(\mathbf{x})$ and the input directions $\mathbf{\hat{s}_1}(\mathbf{x})$ on $z=z_0$ relate the source coordinate $\mathbf{x}$ and the target coordinates $\mathbf{u}(\mathbf{x})$ on $z=z_T$ through a coordinate transformation. This coordinate transformation can be expressed through $\mathbf{\hat{s}_1}(\mathbf{x})$, $z(\mathbf{x})$ and the normalized surface normal vector field $\mathbf{\hat{n}}(\mathbf{x})$.
}
\label{fig:1}
\end{figure}

In \cite{Boe17_2} a mathematical model in form of a MAE for the design of single freeform surfaces for input wavefronts of light sources with zero \'etendue was presented. Thereby a given input wavefront or normalized input ray direction vector field $\mathbf{\hat{s}_1}(\mathbf{x})$ with $\mathbf{x}=(x,y)$ on a plane $z=z_0$, respectively, was assumed (see Fig. \ref{fig:1}).

The derivation of the MAE in \cite{Boe17_2}  was based on the idea that a surface $z(\mathbf{x})$, which is hit by an input ray direction vector field $\mathbf{\hat{s}_1}(\mathbf{x})$, induces a coordinate transformation between the source points $\mathbf{x}$ on $z=z_0$ and the target points $\mathbf{u}(\mathbf{x})$ on the plane $z=z_T$.
By using the well-known ray-tracing equations

\begin{equation}\label{eq:1}
\begin{aligned}
\mathbf{\hat{s}}_2 =n\mathbf{\hat{s}}_1 + \left\{-n\cdot \mathbf{\hat{n}} \cdot \mathbf{\hat{s}}_1 + \sqrt{1- n^2 [1-(\mathbf{\hat{n}} \cdot \mathbf{\hat{s}}_1)^2]} \right\}\mathbf{\hat{n}},\\
\mathbf{\hat{s}}_2 =\mathbf{\hat{s}}_1 -2(\mathbf{\hat{n}}\cdot \mathbf{\hat{s}}_1)\mathbf{\hat{n}}
\end{aligned}
\end{equation}

for refractive and reflective surfaces, which connect the deflected vector field $\mathbf{\hat{s}_2} (\mathbf{x})$ to the input directions $\mathbf{\hat{s}_1}(\mathbf{x})$ and the surface normal vector field $\mathbf{\hat{n}}(\mathbf{x})$ of $z(\mathbf{x})$, the coordinate transformation 

\begin{gather}\label{eq:2}
\mathbf{u}(\mathbf{x})=\mathbf{f}(\mathbf{\hat{s}_1}(\mathbf{x}), z_S(\mathbf{x}), \partial_x z_S(\mathbf{x}), \partial_y z_S(\mathbf{x}) )
\end{gather}

was derived explicitly \cite{Boe17_2}. Thereby  $z_S(\mathbf{x})$ was defined on a scattered $x_S-y_S$ grid through \cite{Boe17_2}

\begin{gather}\label{eq:2b}
\begin{aligned}
z_S(\mathbf{x})\equiv z(\mathbf{x}_S )\\
\mathbf{x}_S   =
\frac{z(\mathbf{x_S})-z_0 }{(\mathbf{\hat{s}_1})_z(\mathbf{x})}
\cdot
\begin{pmatrix}
  (\mathbf{\hat{s}_1})_x (\mathbf{x}) \\
  (\mathbf{\hat{s}_1})_y (\mathbf{x})
\end{pmatrix}
+
\mathbf{x}.
\end{aligned}
\end{gather}

The latter coordinate transformation thereby related the freeform surface intersection points $\mathbf{x_S}$, which are a priori unknown for noncollimated input beams, with the initial coordinates $\mathbf{x}$ on $z=z_0$. The plugging of Eq. (\ref{eq:2}) into the energy conservation equation

\begin{gather}\label{eq:3}
det(\nabla \mathbf{u}(\mathbf{x}))I_T (\mathbf{u}(\mathbf{x}))=I_S (\mathbf{x}),
\end{gather}

led then to the MAE for the unknown surface $z_S (\mathbf{x})$.  This PDE (system) had to be solved by applying the transport boundary conditions $\mathbf{u}(\partial \Omega_S) = \partial \Omega_T$ with $\partial \Omega_S$ and $\partial \Omega_T$ representing the boundary of the source and target distribution, respectively.

\subsection{Generalization to Double Freeform Surfaces}
\label{sec:2.2} 

In the following we demonstrate how the SFD model can be extended to double freeform surfaces for irradiance and phase control for arbitrary input \textit{and} output wavefronts. As it will be seen, in contrast to the SFD, the corresponding PDE system consisting of Eqs. (\ref{eq:2}) and (\ref{eq:3}) will not reduce to a MAE if both wavefronts are nonplanar.

\begin{figure}[!htb]
\begin{center}
\begin{tabular}{c}
\includegraphics[width=\linewidth]{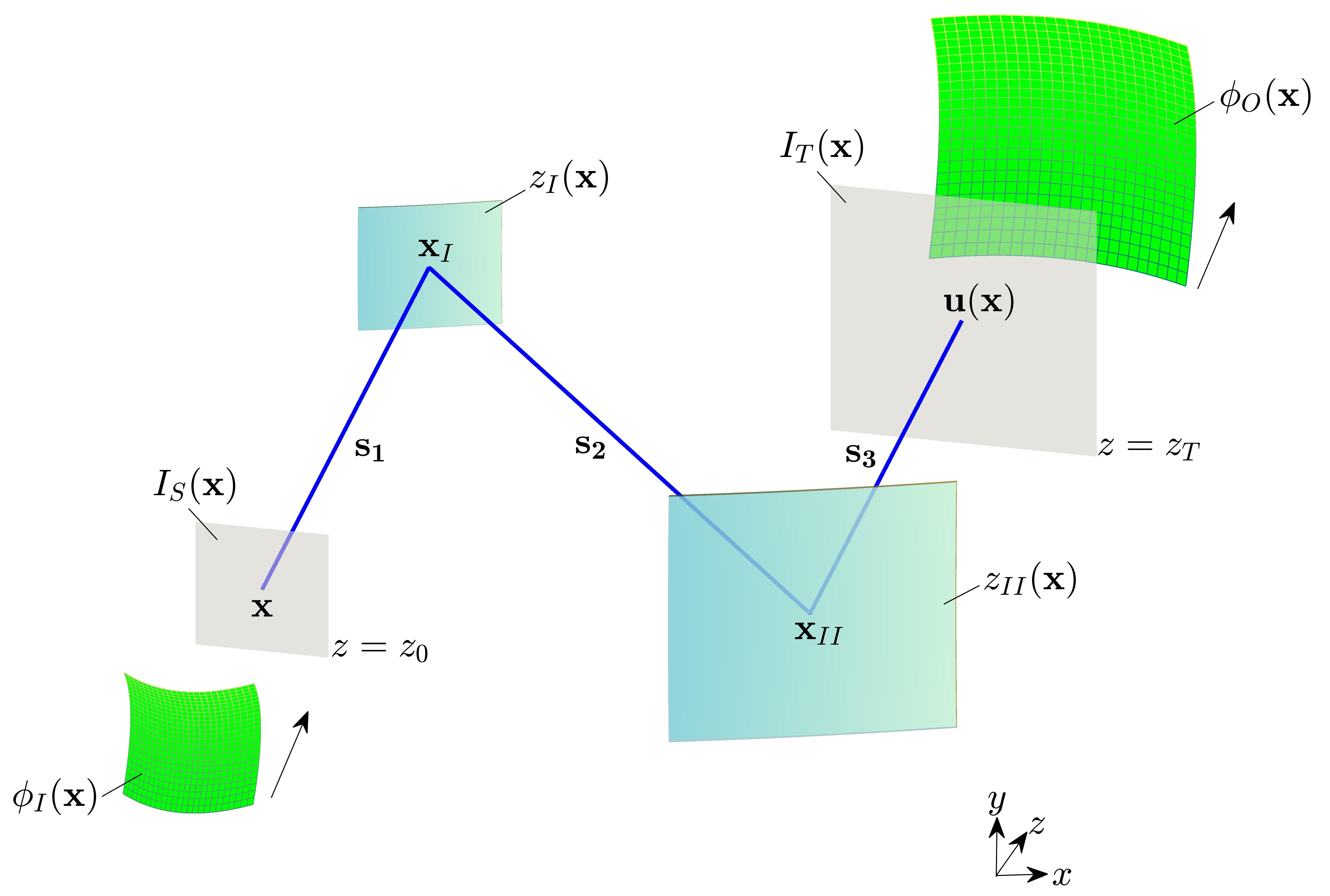}
\end{tabular}
\end{center}
\caption 
{
Geometry of the design problem for the example of double freeform \textit{mirrors}. The predefined input wavefront $\phi_{I}(\mathbf{x})$ defines a normalized input ray direction vector field $\mathbf{\hat{s}_1}(\mathbf{x})$ on $z=z_0$. This input wavefront and the prescribed distribution $I_S (\mathbf{x})$ on $z=z_0$ have to be redistributed by the freeform surfaces $z_I (\mathbf{x})$ and $z_{II}(\mathbf{x})$ to give appropriate target coordinates $\mathbf{u}(\mathbf{x})$ and the required irradiance $I_T (\mathbf{x})$ on $z=z_T$ and output wavefront $\phi_{O}(\mathbf{x})$. For nonplanar wavefronts and unknown surfaces the intermediate coordinates $\mathbf{x}_I$ and $\mathbf{x}_{II}$ are a priori unknown.}
\label{fig:2}
\end{figure}

The geometry of the considered design problem is shown in Fig. \ref{fig:2} for the example of a double \textit{mirror} system, but the equations presented in the following are also valid for double freeform lens systems. The difference is thereby the replacement of the refractive indices for mirrors and the deflected vector field $\mathbf{\hat{s}}_2 (\mathbf{x})$ according to Eq. (\ref{eq:1}).

According to Fig. \ref{fig:2}, the geometry of the problem can be expressed through the vector fields

\begin{gather}\label{eq:4}
\mathbf{s}_1 =
\begin{pmatrix}
  x_{I} - x \\
  y_{I}  - y \\
  z_{I} (\mathbf{x_{I}})-z_0
\end{pmatrix}, 
\ \
\mathbf{s}_3 =
\begin{pmatrix}
  u_x (\mathbf{x}) -x_{II}\\
  u_y (\mathbf{x}) -y_{II}\\
  z_T - z_{II} (\mathbf{x_{II}})
\end{pmatrix},
\nonumber
\\
\mathbf{n_{I}} =
\begin{pmatrix}
  - \partial_{x_{I}} z_{I} (\mathbf{x_{I}}) \\
  - \partial_{y_{I}} z_{I} (\mathbf{x_{I}}) \\
  1
\end{pmatrix},
\ \
\mathbf{n_{II}} =
\begin{pmatrix}
  - \partial_{x_{II}} z_{II} (\mathbf{x_{II}}) \\
  - \partial_{y_{II}} z_{II} (\mathbf{x_{II}}) \\
  1
\end{pmatrix},
\end{gather}

with the surface intersection points

\begin{equation}\label{eq:5}
\begin{aligned}
\mathbf{x_{I}}=
\frac{z_{I,S}(\mathbf{x})-z_0}{(\mathbf{\hat{s}_1})_z (\mathbf{x})}
\cdot
\begin{pmatrix}
  (\mathbf{\hat{s}_1})_x (\mathbf{x}) \\
  (\mathbf{\hat{s}_1})_y (\mathbf{x})
\end{pmatrix}
+
\mathbf{x}
\\
\mathbf{x_{II}}=
\frac{z_{II,S}(\mathbf{x})-z_T}{(\mathbf{\hat{s}_3})_z (\mathbf{u})}
\cdot
\begin{pmatrix}
  (\mathbf{\hat{s}_3})_x (\mathbf{u}) \\
  (\mathbf{\hat{s}_3})_y (\mathbf{u})
\end{pmatrix}
+
\mathbf{u}
\end{aligned}
\end{equation}

with $z_{I,S} (\mathbf{x}) \equiv z_I(\mathbf{x_{I}})$ and $z_{II,S} (\mathbf{x}) \equiv z_{II}(\mathbf{x_{II}})$.

If the input and output wavefronts $\phi_{I} (\mathbf{x})$ and $\phi_{O} (\mathbf{x})$ are given, the corresponding normalized ray directions vector fields $\mathbf{\hat{s}_1}(\mathbf{x})$ and $\mathbf{\hat{s}_3}(\mathbf{x})$ on the source plane $z=z_0$ and target plane $z=z_T$ can directly be calculated from the wavefront gradients.
From these vector fields, the explicit expression of Eq. (\ref{eq:2}) for double freeform surfaces can be derived analogously to \cite{Boe17_2}, which gives

\begin{equation}\label{eq:6}
\begin{aligned}
\mathbf{u} (\mathbf{x})-\mathbf{x}
=
\frac{z_{II,S}(\mathbf{x})-z_{I,S}(\mathbf{x})}{(\mathbf{\hat{s}_2})_z (\mathbf{x})}
\begin{pmatrix}
  (\mathbf{\hat{s}_2})_x (\mathbf{x}) \\
  (\mathbf{\hat{s}_2})_y (\mathbf{x})
\end{pmatrix}
\\
 + 
 \frac{z_{I,S}(\mathbf{x})-z_0}{(\mathbf{\hat{s}_1} )_z (\mathbf{x})}
\begin{pmatrix}
  (\mathbf{\hat{s}_1})_x (\mathbf{x}) \\
  (\mathbf{\hat{s}_1})_y (\mathbf{x})
\end{pmatrix}
-
\frac{z_{II,S}(\mathbf{x})-z_T}{(\mathbf{\hat{s}_3})_z (\mathbf{u})}
\begin{pmatrix}
  (\mathbf{\hat{s}_3})_x (\mathbf{u}) \\
  (\mathbf{\hat{s}_3})_y (\mathbf{u})
\end{pmatrix}.
\end{aligned}
\end{equation}

Hereby $\mathbf{\hat{s}_1} (\mathbf{x})$ and $\mathbf{\hat{s}_3} (\mathbf{x})$ are predefined and the deflected vector field $\mathbf{\hat{s}_2} (\mathbf{x})$ can be expressed by the ray tracing equations in Eq. (\ref{eq:1}) through $\mathbf{\hat{s}_1} (\mathbf{x})$ and $\mathbf{\hat{n}_I} (\mathbf{x})$.

The main difference to the single freeform case 
is therefore the appearance of the $\mathbf{\hat{s}_3} (\mathbf{x})$ term, which depends on $\mathbf{u} (\mathbf{x})$ itself and the coupling to the second freeform surface $z_{II,S}(\mathbf{x})$ in the $\mathbf{\hat{s}_2} (\mathbf{x})$ term.

We want to point out that Eq. (\ref{eq:6}) is valid for two-mirror system, single lens systems with two freeform surfaces and two-lens systems with the target plane \textit{within} the medium of the second lens. Nevertheless, it is straightforeward to generalize Eq. (\ref{eq:6}) to two-lens systems with a finite working distance relative to the exit surface of the second lens (see Appendix B). Also the following concepts can be applied directly without additional difficulties. Alternatively, it is also possible to propagate the predefined target irradiance and output wavefront into the second lens and use the intermediate $I_T (\mathbf{x})$ and $\phi_O (\mathbf{x})$ for the calculations. Depending on the wavefront, this might lead to more complicated boundary shapes of the intermediate $I_T (\mathbf{x})$ and therefore to a more difficult implementation of the transport boundary conditions.

\subsection{OPL Condition}
\label{sec:2.3}

\begin{figure}[!htb]
\begin{center}
\begin{tabular}{c}
\includegraphics[width=\linewidth]{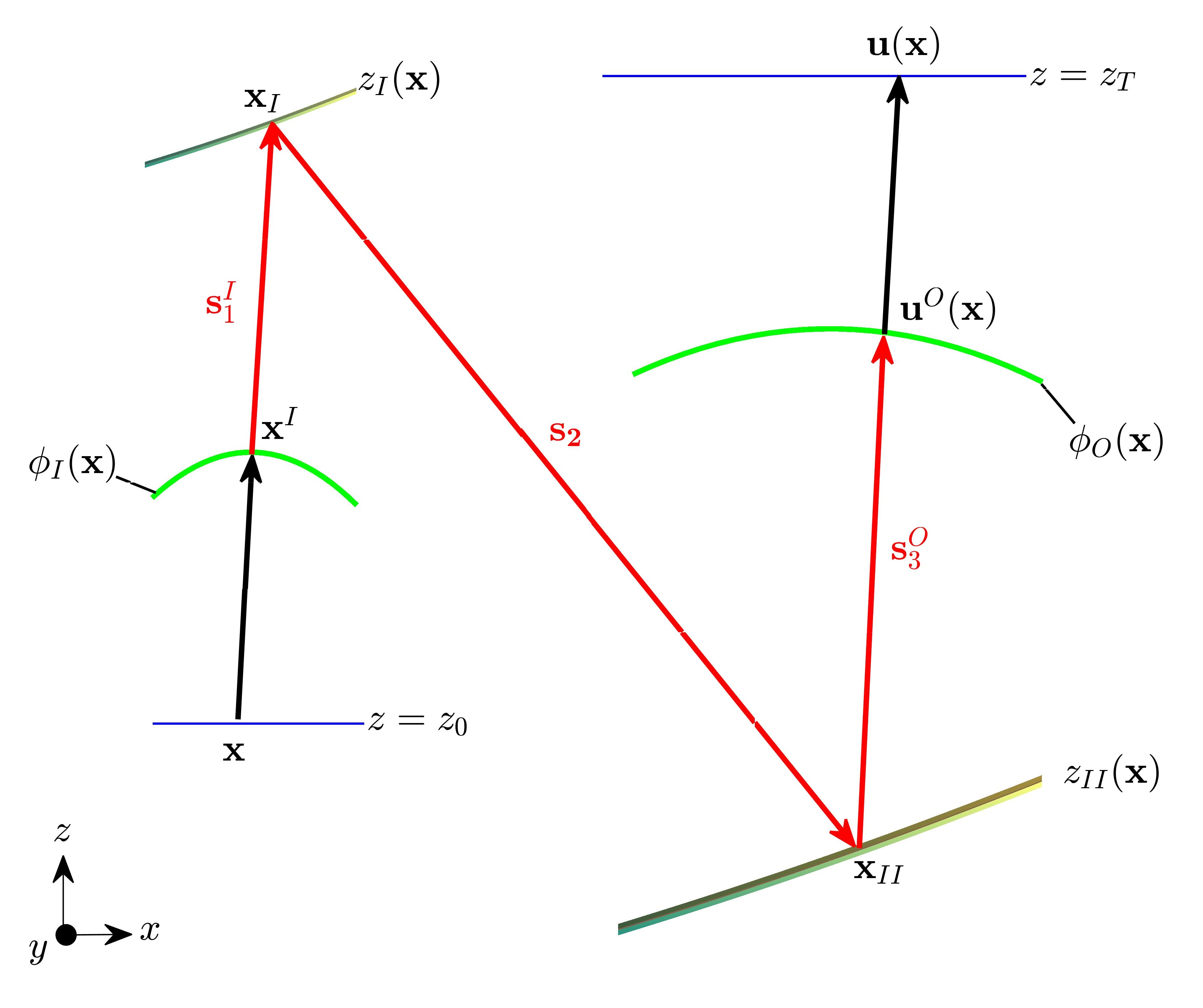}
\end{tabular}
\end{center}
\caption 
{The OPL in Eq. (\ref{eq:7}) is defined through the vector fields $\mathbf{s}_1^{I}$ and $\mathbf{s}_3^{O}$ between the wavefronts and the freeform surfaces. The mapping $\mathbf{u}(\mathbf{x})$ on $z=z_T$ can be related to the mapping coordinates $\mathbf{u}^{O}(\mathbf{x})$ on the wavefront.}
\label{fig:3}
\end{figure}

As pointed out in the previous subsection, additional complications in the design process compared to the SFD arise due to the coupling of the first surface $z_{I,S}(\mathbf{x})$ and the second surface $z_{II,S}(\mathbf{x})$ in Eq. (\ref{eq:6}). This dependency of the PDE system (\ref{eq:3}) and (\ref{eq:6}) on $z_{II,S}(\mathbf{x})$ can be elliminated using the constant OPL condition. By considering 

\begin{equation}\label{eq:7}
OPL=n_1 \cdot |\mathbf{s}_1^{I}|+n_2 \cdot |\mathbf{s}_2|+n_1 \cdot |\mathbf{s}_3^{O}|
\end{equation}

with the vector fields (see Fig. \ref{fig:3})

\begin{gather}\label{eq:8}
\mathbf{s}_1^{I} =
\begin{pmatrix}
  x_{I} - x^{I} \\
  y_{I} - y^{I} \\
  z_{I,S} (\mathbf{x})-\phi_{I}(\mathbf{x}^{I})
\end{pmatrix}, 
\ \ \
\mathbf{s}_3^{O} =
\begin{pmatrix}
  u_x^{O}- x_{II}\\
  u_y^{O}- y_{II} \\
  \phi_{O}(\mathbf{u}^{O}) - z_{II,S} (\mathbf{x})
\end{pmatrix},
\end{gather}

Eq. (\ref{eq:7}) can be solved analytically for $z_{II,S}(\mathbf{x})$. After plugging the solution into Eq. (\ref{eq:6}), the Eqs. (\ref{eq:3}) and (\ref{eq:6}) reduce to a system of three nonlinear PDEs for three unknown functions $u_x (\mathbf{x}) ,u_y (\mathbf{x})$ and $z_{I,S} (\mathbf{x})$.

\subsection{Wavefront Mapping Coordinates and PDE System}
\label{sec:2.4} 

The ellimination of the second surface from the Eq. (\ref{eq:6}) leads to the major difference of the double freeform design compared to the SFD, which is the inherent dependency on the projected mapping coordinates $\mathbf{u}^{O}(\mathbf{x})$ (see Fig. \ref{fig:3}) through the output wavefront $\phi_{O}(\mathbf{u}^{O})$.
The PDE system in Eqs. (\ref{eq:3}) and (\ref{eq:6}) can therefore not be solved directly for $\mathbf{u} (\mathbf{x})$ and $z_{I,S} (\mathbf{x})$. However, since there is a one-to-one correspondence between the mapping $\mathbf{u}(\mathbf{x})$ and $\mathbf{u}^{O}(\mathbf{x})$ through the relations

\begin{gather}\label{eq:9}
\mathbf{u}=
\frac{z_T - \phi_{O}(\mathbf{u}^{O})}{(\mathbf{\hat{s}_3})_z (\mathbf{u}^{O})}
\cdot
\begin{pmatrix}
  (\mathbf{\hat{s}_3})_x (\mathbf{u}^{O}) \\
  (\mathbf{\hat{s}_3})_y (\mathbf{u}^{O})
\end{pmatrix}
+
\mathbf{u}^{O},
\end{gather}

a PDE system for $u_x^{O}(\mathbf{x}), u_y^{O}(\mathbf{x})$ and $z_{I,S} (\mathbf{x})$ can be derived instead. This is done by plugging Eq. (\ref{eq:9}) into Eqs. (\ref{eq:3}) and (\ref{eq:6}). This leaves us with a PDE system of the form

\begin{equation}\label{eq:10}
\begin{aligned}
f (\mathbf{u}^{O},\nabla u_x^{O}, \nabla u_y^{O} ) I_T^{O} (\mathbf{u}^{O})=I_S (\mathbf{x}),
\\
\mathbf{u}^{O} (\mathbf{x})-\mathbf{x} =
\mathbf{f}(z_{I,S},\nabla z_{I,S}, \phi_{O}(\mathbf{u}^{O}),\mathbf{\hat{s}_3} (\mathbf{u}^{O})).
\end{aligned}
\end{equation}

and boundary conditions, which follow directly from $\mathbf{u}(\partial \Omega_S) = \partial \Omega_T$ and Eq. (\ref{eq:9}). Hereby, we redefined $I_T (\mathbf{u})$ through the projected mapping by $I_T^{O} (\mathbf{u}^{O})$. 

We note that we omit to state Eq. (\ref{eq:10}) explicitely due to its lengthiness, but want to emphasize that it can be derived straightforwardly by plugging at first $z_{II,S} (\mathbf{x})$ from the OPL condition into Eq. (\ref{eq:6}) and then Eq. (\ref{eq:9}) into Eqs. (\ref{eq:3}) and (\ref{eq:6}).

In the following section, we will present a possible numerical solving strategy for the PDE system in Eq. (\ref{eq:10}), which generalizes the approach from \cite{Boe17_2} to double freeform surfaces.

\subsection{Numerical Approach}
\label{sec:2.5} 

The first-order PDE system in Eq. (\ref{eq:10}) has a similar structure compared to the PDE system from \cite{Boe17_2}, since two mapping components $u_x^{O}(\mathbf{x})$ and $u_y^{O}(\mathbf{x})$ and the surface $z_{I,S}(\mathbf{x})$ have to be determined simultaneously from three PDEs and the transport boundary conditions.

We therefore discretize $u_x^{O}(\mathbf{x})$ and $u_y^{O}(\mathbf{x})$ and the surface $z_{I,S}(\mathbf{x})$ on an equidistant grid and use first order finite differences for the derivatives at the inner grid points and second order finite differences at the boundary points. This leads to a nonlinear equation system for the unknowns $(u_{x}^{O})_{i;j}, (u_{y}^{O})_{i;j}$ and $(z_{I,S})_{i;j}$ with $i=1,...,N; j=1,...,N$, which we want to solve by standard methods like the trust-region reflective solver from MATLAB 2015b's optimization toolbox. 

Hence, appropriate intial values for $(u_{x}^{O})_{i;j}, (u_{y}^{O})_{i;j}$ and $(z_{I,S})_{i;j}$, which we will declare in the following by the superscript  "$\infty$", are required to ensure a fast convergence of the root finding. Their compuation is done by first calculating the initial map $\mathbf{u}^{\infty}(\mathbf{x})$ from optimal transport \cite{Sul11_1} and then constructing $\mathbf{u}^{O,\infty}(\mathbf{x})$ and $z_{I,S}^{\infty}(\mathbf{x})$ from it. 

The projected map $\mathbf{u}^{O,\infty}(\mathbf{x})$ can be determined by solving the coupled equations in Eq. (\ref{eq:9}) for every value of $\mathbf{u}^{\infty}(\mathbf{x})$. Compared to the SFD this step leads to an additional computional time consumption.

For the given input and output ray direction vector fields $\mathbf{\hat{s}_1} (\mathbf{x})$, $\mathbf{\hat{s}_3} (\mathbf{x})$ and mapping $\mathbf{u}^{\infty}(\mathbf{x})$, the intial surface $z_{I,S}^{\infty}(\mathbf{x})$ can be constructed from a coupled ordinary differential equation (ODE) system (see Appendix A), which is derived directly by inverting the law of refraction/reflection \cite{Rub01_1}. 

The resulting Eqs. (\ref{eq:A1}) and (\ref{eq:A2}) can be solved by fixing the positions of both freeform surfaces through the integration constants $z_{I,S}(\mathbf{x_{I,0}})$ and $z_{II,S}(\mathbf{x_{II,0}})$ and by integrating the ODE system on an arbitrary path on the support of $I_S(\mathbf{x})$ with e.g. MATLAB's ode45 solver. Since the mapping $\mathbf{u}^{\infty}(\mathbf{x})$ is in general not integrable \cite{Boe16_1, Boe17_1}, the  initial surface $z_{I,S}^{\infty}(\mathbf{x})$ will vary with the integration path. Alternatively, the intial surface might be determined from $\mathbf{u}^{\infty}(\mathbf{x})$ by using the method from Feng et al. \cite{Feng13_2}, which was utilised in Ref \cite{Wu16_1}.  

After solving the nonlinear PDE system in Eq. (\ref{eq:10}) for $z_{I,S}(\mathbf{x})$ and $\mathbf{u}^{O}(\mathbf{x})$ with the constructed initial iterate, the second surface $z_{II,S}(\mathbf{x})$ is calculated from the solution of the OPL condition (\ref{eq:7}).

As described in section \ref{sec:2.2}, both surfaces $z_{I,S}(\mathbf{x})$ and $z_{II,S}(\mathbf{x})$ are given on scattered grid points, which are defined by the relations in Eq. (\ref{eq:5}). Depending on the purpose, the surfaces can then be interpolated to the required grid points by scattered data interpolation \cite{MatFE_1}.

\begin{figure}[!htb]
\begin{center}
\begin{tabular}{c}
\includegraphics[width=\linewidth]{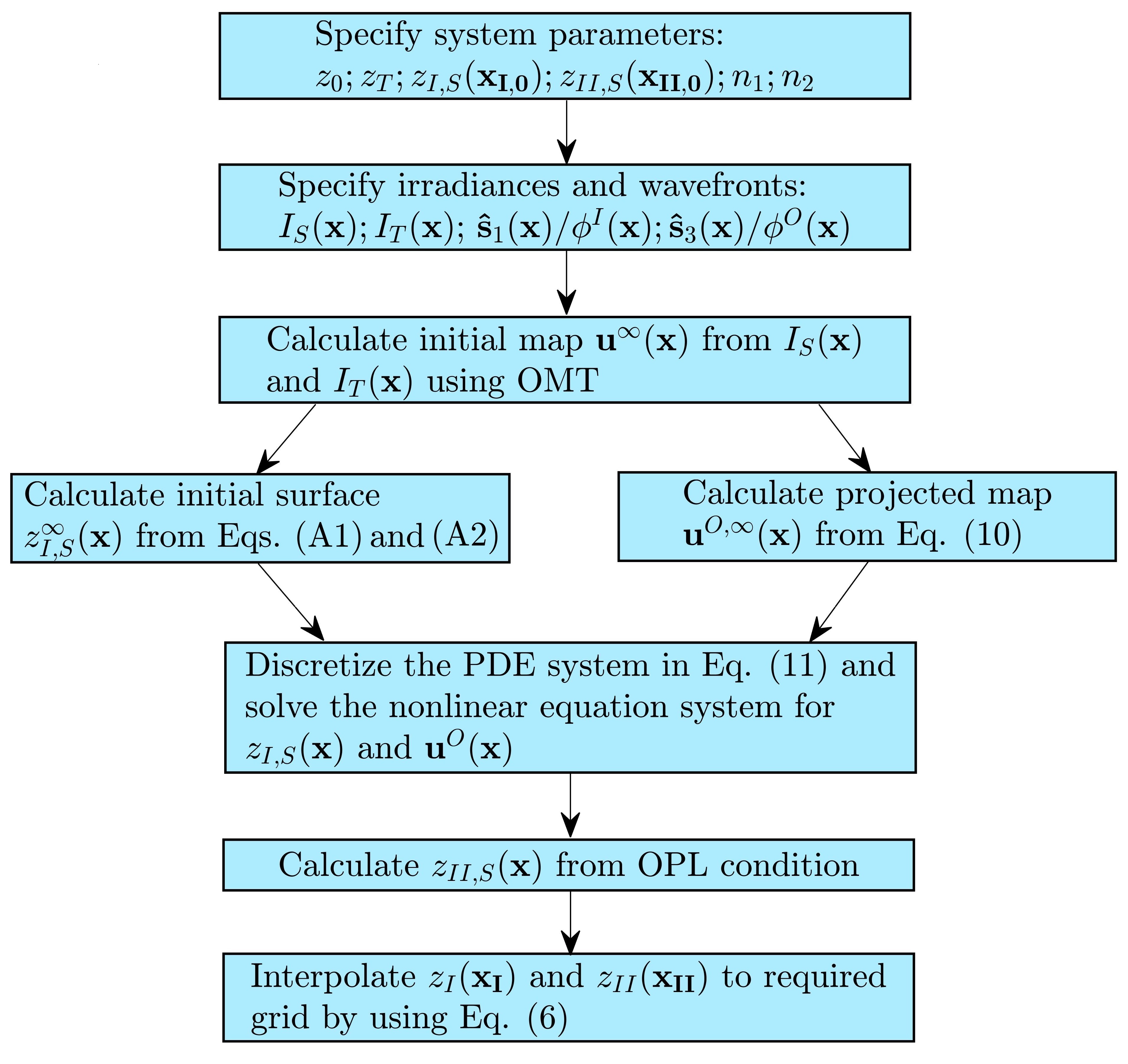}
\end{tabular}
\end{center}
\caption{
Workflow of the double freeform design process.}
\label{fig:4}
\end{figure}

The corresponding workflow of the design process is summarized in Fig. \ref{fig:4}.

\section{Design Examples}
\label{sec:3} 

\begin{figure}[!htb]
\begin{center}
\begin{tabular}{c}
\includegraphics[width=5cm]{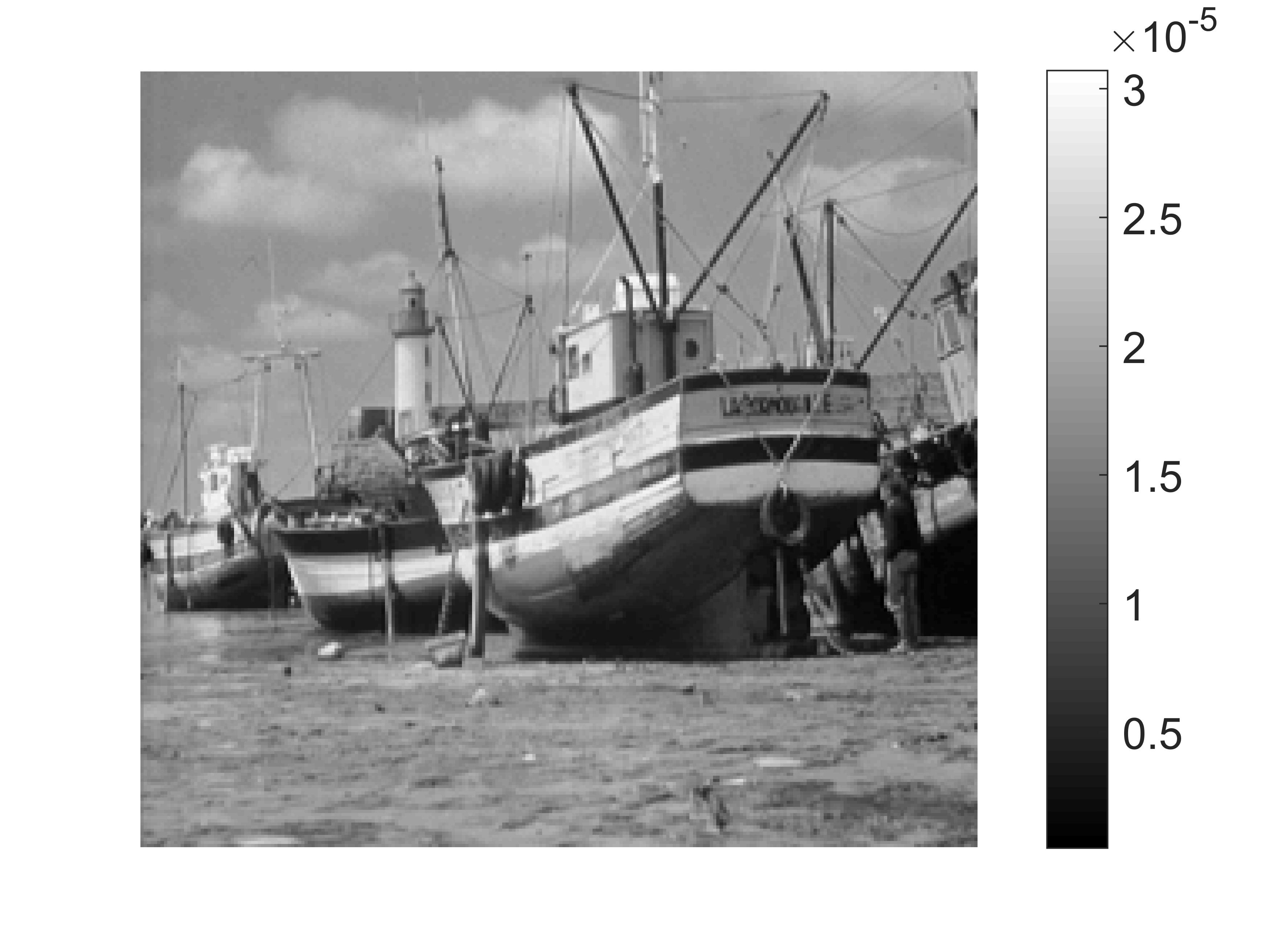}
\end{tabular}
\end{center}
\caption 
{Normalized predefined target distribution ``boat'' with a resolution of $250 \times 250$ pixels.}
\label{fig:5}
\end{figure}

In the following, we want to demonstrate the feasability of the design strategy by applying it to the design of a double mirror and a single lens system with two freeform surfaces. For the examples we use the testimage ``boat'' with a resolution of $250 \times 250$ pixels as the target distribution [Fig. \ref{fig:5}]. To evaluate the quality of the calculated freeform surfaces, we import them into a self-programmed MATLAB ray-tracing toolbox. The irradiances from the ray-tracing simulations are then compared to the predefined irradiance and the optical path difference (OPD) between the predefined wavefronts is calculated to determine the quality of the required phase redistribution. 

To compare the irradiances, the root-mean square $rms_{\Delta_{I_T}}$ of the difference between the predefined and simulated target patterns and the correlation coefficient $corr_{I_T}$ are calculated. The OPD is characterized by the root-mean square $rms_{OPD}$ using a reference wavelength of $550$nm. Additionally, the energy efficiency $\eta$ is calculated. For the ray tracing $200\cdot 10^6$ rays are used. All the calculations are done on an Intel Core i3 at $2 \times 2.4$Ghz with $16$GB RAM.

\begin{figure}[!htb]
\begin{center}
\begin{tabular}{c}
\includegraphics[width=\linewidth]{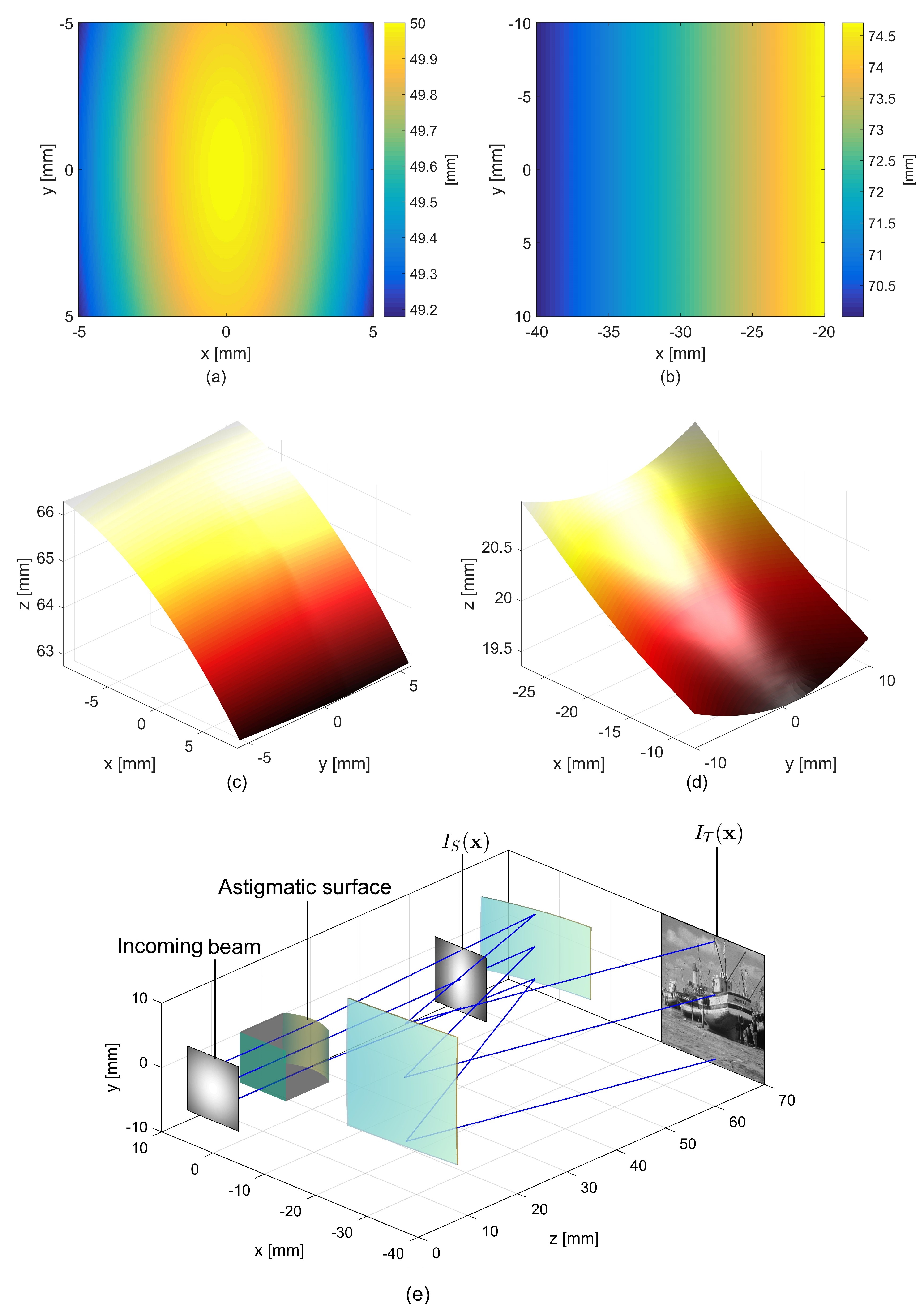}
\end{tabular}
\end{center}
\caption 
{(a) Input wavefront at $z=50$mm produced by the predefined astigmatic surface. (b) Tilted collimated output wavefront at the detector. (c) First freeform mirror. (d) Second freeform mirror. (e) System Layout with two freeform mirrors. An input Gaussian beam with a waist of $w=10$ mm is redistributed by a predefined astigmatic surface to give the input irradiance and ray directions at $z=50$mm. This input distribution $I_S (x,y)$ is redistributed by the double freeform mirror system to give the required output irradiance ``boat'' and a tilted plane wavefront at $z=70$mm.}
\label{fig:6}
\end{figure}

For the mirror example, we use the predefined astigmatic surface $z_{pre}(x,y)$ from \cite{Boe17_2} to generate the input wavefront [Fig. \ref{fig:6}(a)] and irradiance $I_S(\mathbf{x})$ at $z=50$mm from an incoming collimated beam with Gaussian distribution ($w=10$ mm). The goal is to design design two freeform mirrors, which create a tilted collimated output wavefront [Fig. \ref{fig:6}(b)] and a target distribution, which is shifted and scaled relative to the input beam [Fig. \ref{fig:6}(e)]. For the integration constants of the surfaces we choose $z_{I,S,0}=65$mm and $z_{II,S,0}=20$mm. These, together with the predefined wavefront gradients at the boundaries of $I_S(\mathbf{x})$ and $I_T(\mathbf{x})$, determine the positions of the freeform surfaces in space and the extent of the freeform surface in the $x$- and $y$- direction, respectively. The resulting system layout can be seen in Fig. \ref{fig:6}(e).

\begin{figure}[!htb]
\begin{center}
\begin{tabular}{c}
\includegraphics[width=\linewidth]{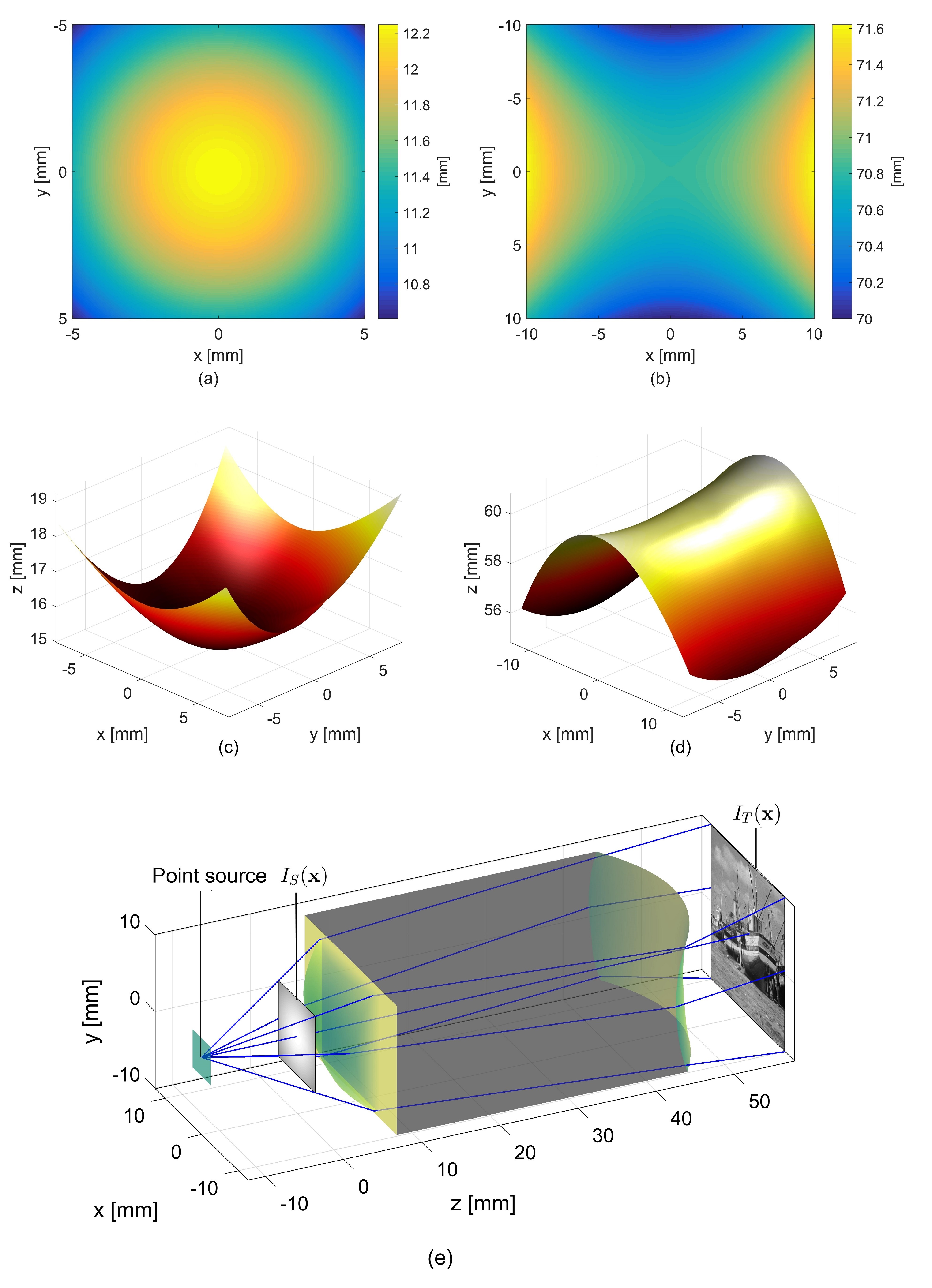}
\end{tabular}
\end{center}
\caption 
{(a) Spherical input wavefront. (b) Astigmatic output wavefront at the detector. (c) First freeform lens surface. (d) Second freeform lens surface. (e) System Layout with two freeform lens surfaces. A point source with a maximum opening angle of $30$deg and a lambertian irradiance distribution $I_S (x,y)$ is redistributed by two freeform lens surfaces to give the required output irradiance ``boat'' and an astigmatic wavefront at $z=70$mm.}
\label{fig:7}
\end{figure}

For the design of the double freeform lens we want to map a point source with a lambertian intensity distribution and an maximum opening angle of $30$deg onto ``boat'' and a predefined astigmatic wavefront [Fig. \ref{fig:7}(b)] at the target plane. The corresponding wavefronts in the source and target plane are presented in Fig. \ref{fig:7}(a) and \ref{fig:7}(b). The integration constants of the freeform surfaces are chosen so that the distance of the point source to $z_{I,S}(0,0)$ is $15$mm and to $z_{II,S}(0,0)$ $60$mm . For the distance of the target plane to $z_{II,S}(0,0)$ $10$mm are used, which leads to the system layout presented in Fig. \ref{fig:7}(e).

After specifying the system geometries by the surface integration constants and the wavefronts and irradiance distributions, the freeform surfaces can be calculated by the design algorithm presented in section \ref{sec:2.5} and Fig. \ref{fig:4}. As shown in Table \ref{table:1}, the major difference in computational time consumption for both examples is the calculation of $\mathbf{u}^{O,\infty}(\mathbf{x})$. Whereas for the tilted collimated output wavefront this can be done by a simple rotation of the coordinate system, the coordinates $\mathbf{u}^{O,\infty}(\mathbf{x})$ for the astigmatic output wavefront are calculated by solving Eq. (\ref{eq:9}) for every target point $\mathbf{u}^{\infty}(\mathbf{x})$ with a nonlinear equation solver.

\begin{table}[!htb]
\centering
\caption{\bf Computational time.}
\begin{tabular}{ccc}
\hline
& Mirror system & Lens system \\
\hline
$\mathbf{u}^{\infty}(\mathbf{x})$		&  $458s$	&  $313s$	\\
$\mathbf{u}^{O,\infty}(\mathbf{x})$		&  $< 1s$ 	&  $260s$	\\
$z_{I,S}^{\infty}(\mathbf{x})$			&  $269s$ &  $251s$	\\
$z_{I,S}(\mathbf{x})$					&  $258s$ &  $234s$	\\
\hline
\end{tabular}
  \label{table:1}
\end{table}

To calculate the intial surfaces MATLABs ode45 solver is used with tolerances of $10^{-8}$. Compared to the SFD in \cite{Boe17_2}, despite the lower tolerances of $10^{-8}$, the compuational time of $z_{I,S}^{\infty}(\mathbf{x})$ is significantly larger, which is due to the coupling to the second freeform surface in Eq. (\ref{eq:A2}). A decoupling of the differential equations by the OPL condition might therefore be helpful to reduce the computational time.

Following the calculation of the initial iterate, the root finding of Eq. (\ref{eq:10}) is done with MATLABs \textit{fsolve()} function, which leads to computational times slightly higher than the SFD \cite{Boe17_2} resulting from increased complexity of Eqs. (\ref{eq:2}) and (\ref{eq:3}), respectively, as discussed in section \ref{sec:2}. The noninterpolated final surfaces are shown in Fig. \ref{fig:6}(c) and Fig. \ref{fig:6}(d) for the double mirror system and in Fig. \ref{fig:7}(c) and Fig. \ref{fig:7}(d) for the double freeform lens.

\begin{figure}[!htb]
\begin{center}
\begin{tabular}{c}
\includegraphics[width=\linewidth]{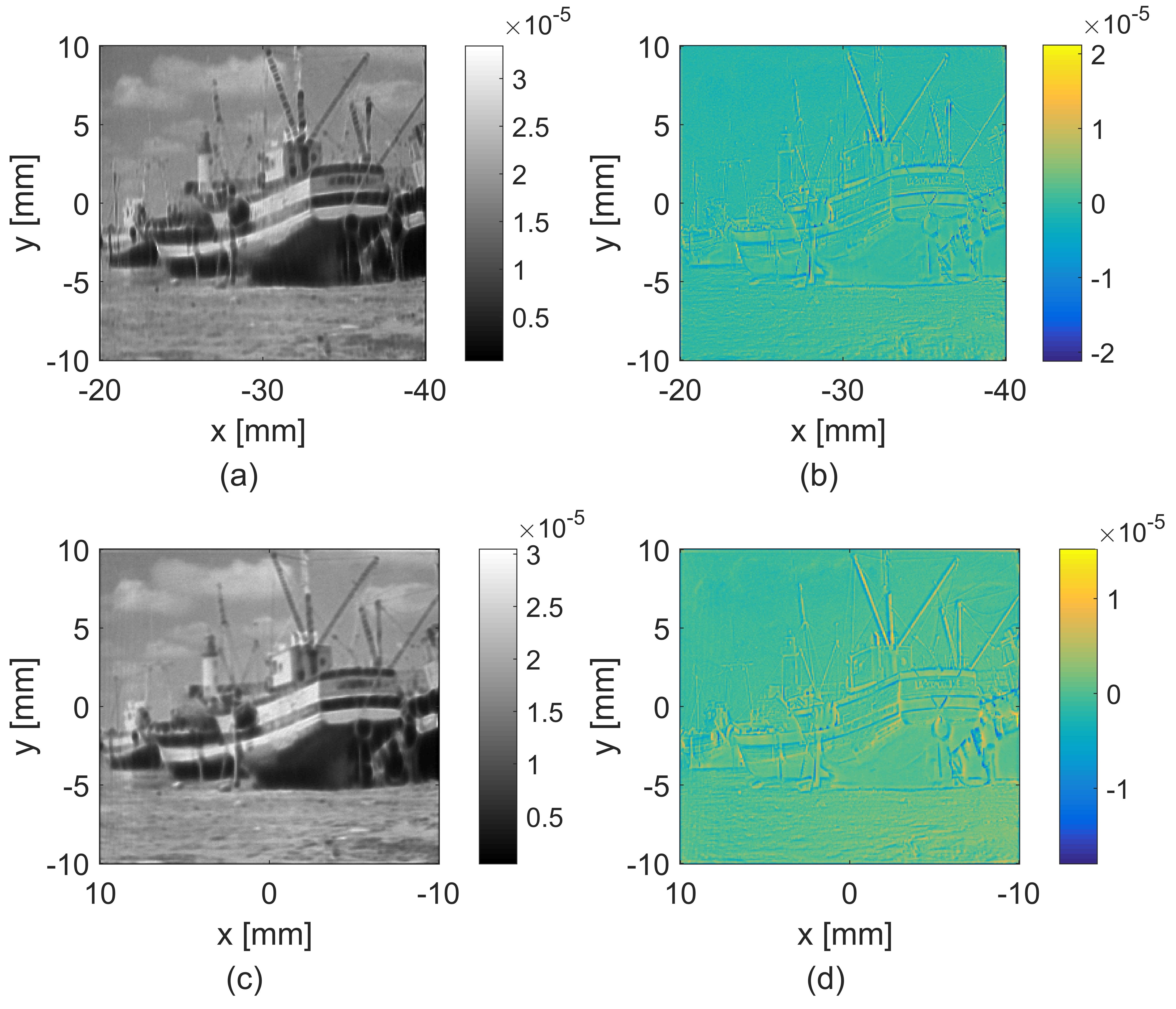}
\end{tabular}
\end{center}
\caption 
{(a) Simulated irradiance for the double freeform mirror. (b) Difference between predefined and simulated irradiance for the double freeform mirror. (c)  Simulated irradiance for the double freeform lens.  (d) Difference between predefined and simulated irradiance for the double freeform lens.}
\label{fig:8}
\end{figure}

\begin{figure}[!htb]
\begin{center}
\begin{tabular}{c}
\includegraphics[width=\linewidth]{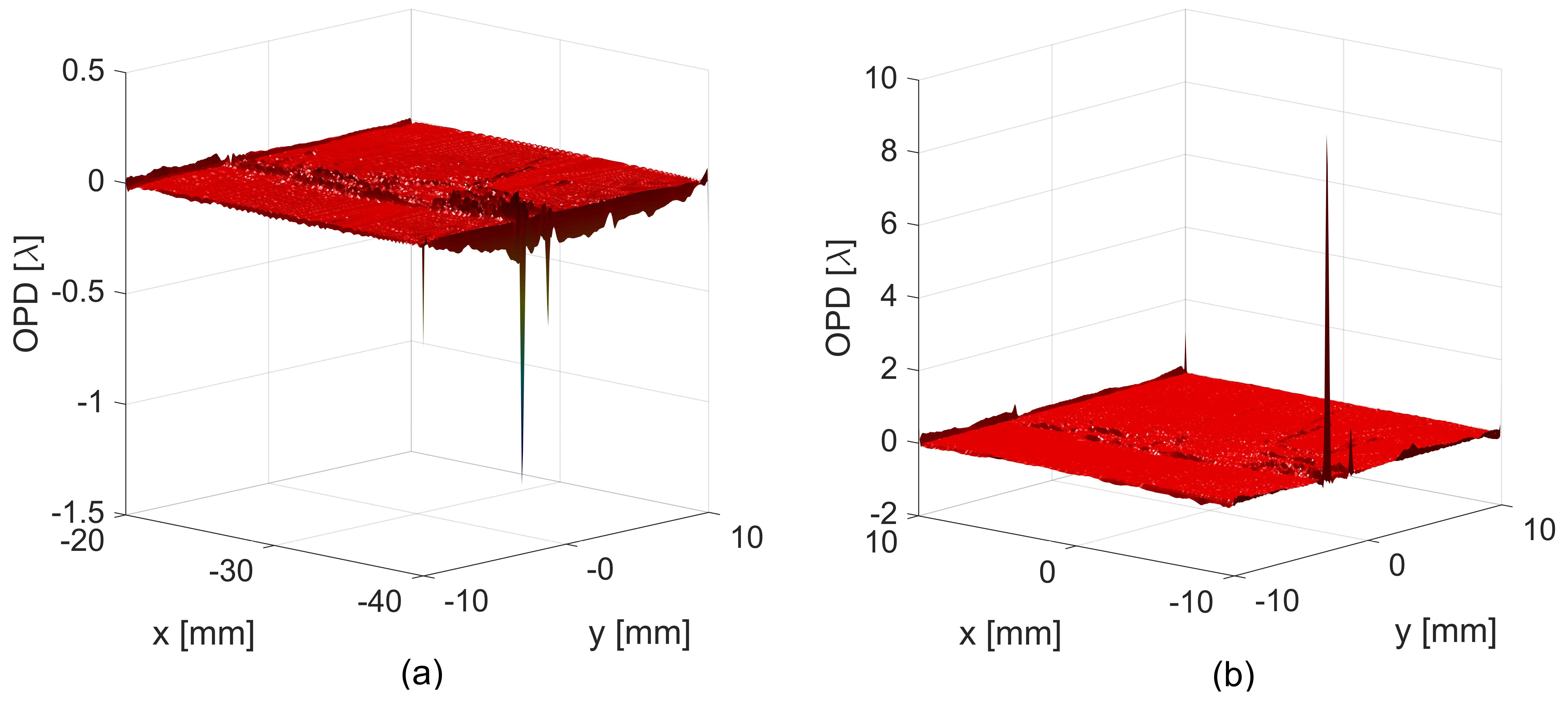}
\end{tabular}
\end{center}
\caption 
{(a) OPD between predefined wavefronts for the double freeform mirror ($rms_{OPD}=0.0154 \lambda$). (b) OPD between predefined wavefronts for the double freeform lens ($rms_{OPD}=0.0670 \lambda$). The OPL deviations occur mainly at the boundary due to the interpolation/extrapolation of the freeform surfaces.}
\label{fig:9}
\end{figure}

After the interpolation and extrapolation of the freeform surfaces onto an equidistant, rectangular grid (see Figs. \ref{fig:6}(e) and \ref{fig:7}(e)), their quality is characterized by a ray-tracing simulation. The results from the ray tracing are shown in Figs. \ref{fig:8} and \ref{fig:9} and Table \ref{table:2}. As it was observed in previous publications \cite{Boe16_1, Boe17_1, Boe17_2}, deviations between the predefined and the simulated irradiance arise mainly at strong gradients in the irradiance distribution [Fig. \ref{fig:8}(b) and Fig. \ref{fig:8}(d)].

The OPDs between the predefined wavefronts [Fig. \ref{fig:9}] and the calculated rms values of $0.0154 \lambda$ (mirror system) and $0.0670 \lambda$ (lens system) show satisfying uniformity beyond the diffraction limit. Major contributions to the $rms_{OPD}$ are thereby due the boundary interpolation/extrapolation of the freeform surfaces onto the equidistant, rectangular grid (see Fig. \ref{fig:9}).

The simulation results for the irradiances and OPDs show the capabilities of the presented design approach for complex illumination design problems.

\begin{table}[!htb]
\centering
\caption{\bf Comparison of $\Delta I_T$ and $OPD$ for the freeform systems.}
\begin{tabular}{ccc}
\hline
 & Mirror System & Lens system \\
\hline
$rms_{\Delta I_T}$	& $2.369 \cdot 10^{-6}$		& $2.0179 \cdot 10^{-6}$	\\
$corr_{I_T}$		& $0.9127$					& $0.9349$			\\
$\eta$				& $99.77$\%					& $99.70$\%			\\
$rms_{OPD}$			& $0.0154 \lambda$ 			& $0.0670 \lambda$	\\
\hline
\end{tabular}
  \label{table:2}
\end{table}

\section{Conclusion}

A mathematical model for the design of double freeform surface systems for irradiance and phase control with arbitrary ideal input and output wavefronts was introduced. The model was derived by expressing the coordinates at the target plane in terms of both freeform surfaces and their surface gradients, by elliminating the second surface from the PDE system with the OPL condition and by using the projected ray-mapping coordinates on the output wavefront. The PDE system consists of three coupled nonlinear PDEs for the first freeform surface and the projected mapping coordinates. A numerical solving strategy was proposed, which was implemented for irradiance distributions on square apertures and tested by applying it to the design of a double freeform mirror and lens system. 

Since the model is neither restricted to planar input or output wavefronts nor to the paraxial regime, it opens up new possibilities in applications of freeform surfaces in beam shaping and illumination design. E.g. the design method can be applied \textit{directly} without additional implementation effort to the design problem of calculating a double freeform surface system for two predefined target irradiances, which was considered in \cite{Feng17_1}. As it was shown in \cite{Rub04_1}, the required output wavefront $\phi_{O}(\mathbf{x})$ between the target distributions can be calculated from optimal mass transport with a quadratic cost function in a paraxial regime with e.g. the method by Sulman et al. \cite{Sul11_1}. The first target irradiance and $\phi_{O}(\mathbf{x})$ serve then as the input for the design algorithm [Fig. \ref{fig:4}]. Since the wavefront calculation itself introduces additional numerical errors, which are independent from the freeform design method, we will focus on the presentation and in-depth discussion of such a design example in our future work.

Similar to \cite{Boe17_2}, the introduction of the model defined by Eq. (\ref{eq:10}), combined with the transport boundary condition demands the development and application of numerical algorithms for more general boundary shapes, which is one of our future goals.

\section*{Appendix A}
\label{sec:Appendix A}
\setcounter{equation}{0} \renewcommand{\theequation}{A.\arabic{equation}} 
\subsection{Initial Surface Construction}
\label{sec:Appendix A1}

For a given ray mapping $\mathbf{u}(\mathbf{x})$ and ray directions $\mathbf{\hat{s}}_1 (\mathbf{x})$ and $\mathbf{\hat{s}}_3 (\mathbf{x})$, the intial surfaces can be calculated from a system of coupled ODEs \cite{Rub01_1}. The system can be derived straightforward from the law of refraction/reflection $\mathbf{n}= n_1 \mathbf{\hat{s}}_1 - n_2 \mathbf{\hat{s}}_2$ and the coordinate transformations in Eq. (\ref{eq:5}), which gives

\begin{equation}\label{eq:A1}
\begin{aligned}
\partial_{x}z_{I,S}^{\infty}(\mathbf{x})
=
\frac{n_{I,x}(\mathbf{x})+h_{I,x}(\mathbf{x})[z_{I,S}^{\infty}(\mathbf{x}) -z_0 ]}{1-n_{I,x} (\mathbf{x})\frac{(\mathbf{\hat{s}}_1 )_x (\mathbf{x})}{(\mathbf{\hat{s}}_1 )_z (\mathbf{x})}- n_{I,y} (\mathbf{x})\frac{(\mathbf{\hat{s}}_1 )_y (\mathbf{x})}{(\mathbf{\hat{s}}_1 )_z (\mathbf{x})} },
\\
\partial_{y}z_{I,S}^{\infty}(\mathbf{x})
=
\frac{n_{I,y}(\mathbf{x})+h_{I,y}(\mathbf{x})[z_{I,S}^{\infty}(\mathbf{x})-z_0 ] }{1-n_{I,x} (\mathbf{x})\frac{(\mathbf{\hat{s}}_1 )_x (\mathbf{x})}{(\mathbf{\hat{s}}_1 )_z (\mathbf{x})}- n_{I,y} (\mathbf{x})\frac{(\mathbf{\hat{s}}_1 )_y (\mathbf{x})}{(\mathbf{\hat{s}}_1 )_z (\mathbf{x})} }
\end{aligned}
\end{equation}

and

\begin{equation}\label{eq:A2}
\begin{aligned}
\partial_{x}z_{II,S}^{\infty}(\mathbf{x})
=
\\
\frac{
n_{II,x}(\mathbf{x})\partial_x u_x +n_{II,y}(\mathbf{x})\partial_x u_y +
h_{II,x}(\mathbf{x})[z_{II,S}^{\infty}(\mathbf{x})-z_T ] }{1-n_{II,x}(\mathbf{x}) \frac{(\mathbf{\hat{s}}_3 )_x (\mathbf{u} )}{(\mathbf{\hat{s}}_3 )_z (\mathbf{u} )}- n_{II,y}(\mathbf{x})\frac{(\mathbf{\hat{s}}_3 )_y (\mathbf{u} )}{(\mathbf{\hat{s}}_3 )_z (\mathbf{u} )} },
\\
\partial_{y}z_{II,S}^{\infty}(\mathbf{x})
=
\\
\frac{
n_{II,x}(\mathbf{x})\partial_y u_x +n_{II,y}(\mathbf{x})\partial_y u_y +
h_{II,y}(\mathbf{x})[z_{II,S}^{\infty}(\mathbf{x})-z_T ] }{1-n_{II,x}(\mathbf{x}) \frac{(\mathbf{\hat{s}}_3 )_x (\mathbf{u})}{(\mathbf{\hat{s}}_3 )_z (\mathbf{u})}- n_{II,y}(\mathbf{x})\frac{(\mathbf{\hat{s}}_3 )_y (\mathbf{u})}{(\mathbf{\hat{s}}_3 )_z (\mathbf{u})} }
.
\end{aligned}
\end{equation}

The coefficients are thereby defined by 

\begin{equation}\label{eq:A3}
\begin{aligned}
h_{I,x}(\mathbf{x})
=
n_{I,x} (\mathbf{x})\partial_x \left(\frac{(\mathbf{\hat{s}}_1 )_x (\mathbf{x})}{(\mathbf{\hat{s}}_1 )_z (\mathbf{x})}\right)+n_{I,y} (\mathbf{x})\partial_x \left(\frac{(\mathbf{\hat{s}}_1 )_y (\mathbf{x})}{(\mathbf{\hat{s}}_1 )_z (\mathbf{x})}\right),
\\
h_{I,y}(\mathbf{x})
=
n_{I,x} (\mathbf{x})\partial_y \left(\frac{(\mathbf{\hat{s}}_1 )_x (\mathbf{x})}{(\mathbf{\hat{s}}_1 )_z (\mathbf{x})}\right)+n_{I,y} (\mathbf{x})\partial_y \left(\frac{(\mathbf{\hat{s}}_1 )_y (\mathbf{x})}{(\mathbf{\hat{s}}_1 )_z (\mathbf{x})}\right),
\\
h_{II,x}(\mathbf{x})
=
n_{II,x}(\mathbf{x})\partial_x \left(\frac{(\mathbf{\hat{s}}_3 )_x (\mathbf{u} )}{(\mathbf{\hat{s}}_3 )_z (\mathbf{u})}\right)+n_{II,y}(\mathbf{x}) \partial_x \left(\frac{(\mathbf{\hat{s}}_3 )_y  (\mathbf{u} )}{(\mathbf{\hat{s}}_3 )_z  (\mathbf{u} )}\right),
\\
h_{II,y}(\mathbf{x})
=
n_{II,x}(\mathbf{x}) \partial_y \left(\frac{(\mathbf{\hat{s}}_3 )_x (\mathbf{u})}{(\mathbf{\hat{s}}_3 )_z (\mathbf{u})}\right)+n_{II,y}(\mathbf{x})\partial_y \left(\frac{(\mathbf{\hat{s}}_3 )_y (\mathbf{u})}{(\mathbf{\hat{s}}_3 )_z (\mathbf{u})}\right), 
\\
 n_{I,x} (\mathbf{x})
 \equiv
-\frac{n_1 (\mathbf{\hat{s}}_1 )_x (\mathbf{x})  -n_2 (\mathbf{\hat{s}}_2)_x (\mathbf{x})}{n_1 (\mathbf{\hat{s}}_1 )_z (\mathbf{x}) -n_2 (\mathbf{\hat{s}}_2 )_z (\mathbf{x})},
\\
 n_{I,y} (\mathbf{x})
\equiv
-\frac{n_1 (\mathbf{\hat{s}}_1 )_y (\mathbf{x}) -n_2 (\mathbf{\hat{s}}_2  )_y (\mathbf{x})}{n_1 (\mathbf{\hat{s}}_1)_z (\mathbf{x}) -n_2 (\mathbf{\hat{s}}_2 )_z (\mathbf{x})},
\\
 n_{II,x} (\mathbf{x})
 \equiv
-\frac{n_1 (\mathbf{\hat{s}}_3 )_x (\mathbf{u})-n_2 (\mathbf{\hat{s}}_2 )_x (\mathbf{x}) }{ n_1 (\mathbf{\hat{s}}_3 )_z (\mathbf{u})-n_2 (\mathbf{\hat{s}}_2)_z (\mathbf{x})},
\\
 n_{II,y} (\mathbf{x})
\equiv
-\frac{ n_1 (\mathbf{\hat{s}}_3 )_y (\mathbf{u}) -n_2 (\mathbf{\hat{s}}_2 )_y (\mathbf{x})}{ n_1 (\mathbf{\hat{s}}_3 )_z  (\mathbf{u}) -n_2 (\mathbf{\hat{s}}_2)_z (\mathbf{x})}.
\end{aligned}
\end{equation}

\subsection{Double Lens with two Freeform Surfaces}
\label{sec:Appendix A2}
\setcounter{equation}{0} \renewcommand{\theequation}{B.\arabic{equation}} 
For double lens systems with two freeform surfaces and a plane exit surface of the second lens, Eq. (\ref{eq:6}) is replaced by

\begin{equation}\label{eq:A4}
\begin{aligned}
\mathbf{u} (\mathbf{x})-\mathbf{x}
=
\frac{z_{II,S}(\mathbf{x})-z_{I,S}(\mathbf{x})}{(\mathbf{\hat{s}_2})_z (\mathbf{x})}
\begin{pmatrix}
  (\mathbf{\hat{s}_2})_x (\mathbf{x}) \\
  (\mathbf{\hat{s}_2})_y (\mathbf{x})
\end{pmatrix}
\\
 + 
 \frac{z_{I,S}(\mathbf{x})-z_0}{(\mathbf{\hat{s}_1} )_z (\mathbf{x})}
\begin{pmatrix}
  (\mathbf{\hat{s}_1})_x (\mathbf{x}) \\
  (\mathbf{\hat{s}_1})_y (\mathbf{x})
\end{pmatrix}
\\
-
\frac{z_{II,S}(\mathbf{x})-z_{M}}{(\mathbf{\hat{s}_3}^{M})_z (\mathbf{u}^{M})}
\begin{pmatrix}
  (\mathbf{\hat{s}_3}^{M})_x (\mathbf{u}^{M}) \\
  (\mathbf{\hat{s}_3}^{M})_y (\mathbf{u}^{M})
\end{pmatrix}
\\
+
\frac{z_T-z_{M}}{(\mathbf{\hat{s}_3})_z (\mathbf{u} )}
\begin{pmatrix}
  (\mathbf{\hat{s}_3})_x (\mathbf{u}) \\
  (\mathbf{\hat{s}_3})_y (\mathbf{u} )
\end{pmatrix}
\end{aligned}
.
\end{equation}

Hereby $\mathbf{u}^{M}(\mathbf{x})$ represents the projection of the mapping $\mathbf{u}(\mathbf{x})$ onto the exit surface plane $z=z_{M}$. The vector field $\mathbf{s}^M_3 (\mathbf{u}^{M} )$ is defined in the medium between the freeform surface $z_{II,S}(\mathbf{x})$ and the plane $z=z_{M}$ and can be expressed by the law of refraction through

\begin{equation}\label{eq:A5}
\begin{aligned}
(\mathbf{\hat{s}_3})_x (\mathbf{u} )=
n(\mathbf{\hat{s}_3}^{M})_x (\mathbf{u}^{M} )
\\
(\mathbf{\hat{s}_3})_y (\mathbf{u} )=
n(\mathbf{\hat{s}_3}^{M})_y (\mathbf{u}^{M} )
\\
(\mathbf{\hat{s}_3})_z (\mathbf{u} )=
\sqrt{1-n^2\{1-[(\mathbf{\hat{s}_3}^{M})_z (\mathbf{u}^{M} )]^2\}}
\end{aligned}
.
\end{equation}

After determining the intermediate output wavefront $\phi_O^M (\mathbf{x})$ in the medium from $\phi_O (\mathbf{x})$ at $z=z_T$, the second surface $z_{II,S}(\mathbf{x})$ can be derived from Eq. (\ref{eq:8}) and plugged into  Eq. (\ref{eq:A4}).
 
There are two options to express Eqs. (\ref{eq:A4}) and (\ref{eq:3}) through the wavefront mapping $\mathbf{u}^{O}(\mathbf{x})$.
The first is by replacing Eq. \ref{eq:9} by

\begin{gather}\label{eq:A6}
\mathbf{u}=
\left\{  
n
\frac{z_T - z_M}{(\mathbf{\hat{s}_3})_z (\mathbf{u})}
+
\frac{z_M - \phi^M_{O}(\mathbf{u}^{O})}{(\mathbf{\hat{s}_3}^M)_z (\mathbf{u}^{O})}
  \right\}
\begin{pmatrix}
  (\mathbf{\hat{s}_3}^M)_x (\mathbf{u}^{O}) \\
  (\mathbf{\hat{s}_3}^M)_y (\mathbf{u}^{O})
\end{pmatrix}
+
\mathbf{u}^{O}.
\end{gather}

Here the mapping $\mathbf{u}(\mathbf{x})$ is expressed through the projected coordinates $\mathbf{u}^O(\mathbf{x})$ onto $\phi_O^M (\mathbf{x})$. Plugging Eq. (\ref{eq:A6}) into Eq. (\ref{eq:A4}) and (\ref{eq:3}) and keeping in mind that $\mathbf{\hat{s}_3}^{M} (\mathbf{u}^{M} ) \equiv \mathbf{\hat{s}_3}^{M} (\mathbf{u}^{O} )$ leads to the analoguous of Eq. (\ref{eq:10}) for double lenses with a plane exit surface.

The second option is to use Eq. (\ref{eq:9}), to solve the OPL condition

\begin{equation}\label{eq:A7}
OPL=n_1 \cdot |\mathbf{s}_1^{I}|+n_2 \cdot |\mathbf{s}_2|+n_1 \cdot |\mathbf{s}_3^{M}|+n_2 \cdot |\mathbf{s}_3^{O}|
\end{equation}

for $z_{II,S}(\mathbf{x})$ and to replace the intermediate coordinate $\mathbf{u}^{M}(\mathbf{x})$ by

\begin{gather}\label{eq:A8}
\mathbf{u}^M=
\frac{z_M - \phi_{O}(\mathbf{u}^{O})}{(\mathbf{\hat{s}_3})_z (\mathbf{u}^{O})}
\cdot
\begin{pmatrix}
  (\mathbf{\hat{s}_3})_x (\mathbf{u}^{O}) \\
  (\mathbf{\hat{s}_3})_y (\mathbf{u}^{O})
\end{pmatrix}
+
\mathbf{u}^{O},
\end{gather}

which leads also to a PDE system like Eq. (\ref{eq:10}) for $\mathbf{u}^{O}(\mathbf{x})$ and $z_{I,S}(\mathbf{x})$.

\section*{Funding}

Bundesministerium f\"ur Bildung und Forschung (BMBF) (FKZ: 031PT609X, FKZ:O3WKCK1D)

\section*{Acknowledgments}

The authors want to thank Norman G. Worku for providing them with the MatLightTracer.


\end{document}